\documentstyle[12pt]{article}
\author{Yu. M. Zinoviev
        \thanks{E-mail address: ZINOVIEV@MX.IHEP.SU} \\
        {\it Institute for High Energy Physics} \\
        {\it Protvino, Moscow Region, 142284, Russia}}
\title{Dual versions of $N=2$ supergravity and \\
       spontaneous supersymmetry breaking}
\date{}
\begin{document}

\maketitle
\thispagestyle{empty}

\begin{abstract}
In this paper, using a model of $N=2$ supergravity - vector multiplets
interaction with the scalar field geometry $SU(1,m)/SU(m)\otimes U(1)$ as
an
example, we show that even when the geometry is fixed one can have a
whole family of the Lagrangians that differ by the vector field
duality transformations. As a byproduct, for this geometry we have
constructed a model of $(m-1)$ vector multiplets interacting with the
hidden sector admitting spontaneous supersymmetry breaking with two
arbitrary scales and without a cosmological term.
\end{abstract}

\newpage
\setcounter{page}{1}

\section{Introduction}

   Recently there was essential progress in understanding of general
coupling of $N=2$ supergravity to vector multiplets (as well as
hypermultiplets) and of geometrical aspects in such models
\cite{D'A91}. As is well known (e.g. \cite{Cre85}) not all of the
isometries of the manifold spanned by scalar fields turn out to be
the symmetry of the Lagrangian, part of them (containing vector fields
duality transformations) being a symmetry of the equations of motion
only. First of all this leads to the existence of different (we will
call them dual) versions of supergravity - vector multiplets
interaction with the same geometry in which part of the vector fields
are replaced by axial vector ones. For example, there exist two
different models for one vector multiplets with the scalar field
geometry $SU(1,1)/U(1)$. One of them is so called minimal model
\cite{Wit84a} when the Lagrangian is invariant under non-compact
subgroup $O(1,1)$. Another one appeared as a part of the hidden sector
in our search \cite{Zin86} for mechanism of spontaneous supersymmetry
breaking with two arbitrary scales (including partial super-Higgs
effect) without a cosmological term. In this model two vector fields
(graviphoton and the matter one) enter the Lagrangian and the
supertransformation laws through the complex combinations $A_\mu  \pm
\gamma _5
B_\mu $, the Lagrangian being invariant under the compact
$U(1)$-subgroup.

   The aim of this paper is to show that the invariance of the
equations of motion under duality transformations leads to the
existence of the whole family of the Lagrangians with the same scalar
field geometry. At first we have constructed the general coupling of
one vector multiplet to $N=2$ supergravity with the geometry
$SU(1,1)/U(1)$ which can smoothly interpolates between two models
mentioned above. Then we have managed to generalize this model to the
case of arbitrary number of vector multiplets, the scalar field
geometry being $SU(1,m)/SU(m)\otimes U(1)$. As a partial but physically a
most interesting case we obtained the model for $(m-1)$ vector
multiplets interacting with our hidden sector. It turns out that
in this model one really can have spontaneous supersymmetry breaking
without a cosmological term.

\section{Dual versions of $N=2$ supergravity}

   Let us use as an example the well known model of the $N=2$
supergravity --- vector multiplets interaction with the scalar field
geometry $SU(1,m)/SU(m)\otimes U(1)$. The most simple way to describe this
model is to introduce, apart from the graviton $e_{\mu r}$ and gravitini
$\Psi _{\mu i}$, $i=1,2$, $m+1$ vector multiplets $\{A_\mu {}^A, \Omega
_{iA}, z^A\}$,
$A=0,1,...m$. The following constraints correspond to the model with
required geometry:
\begin{equation}
 \bar{z}_A \cdot  z^A = - 2 \qquad z^A \cdot  \Omega _{iA} = 0 \label{e1}
\end{equation}
Note that we use the $\gamma $-matrix representation in which majorana
spinors are real, so in all expressions with spinors the matrix $\gamma
_5$
will play a role of the imaginary unit $i$, e.g., $z^A \cdot  \Omega _{iA}
= (x^A
+ \gamma _5 y^A) \cdot  \Omega _{iA}$. In this, the theory has local axial
$U(1)$
invariance with composite gauge field $U_\mu  = (\bar{z}_A \partial _\mu
z^A)$. The
corresponding covariant derivatives look like, e.g.,
\begin{eqnarray}
 {\cal D}_\mu  z^A &=& \partial _\mu  z^A + \frac{1}{2} (\bar{z} \partial
_\mu  z) z^A, \qquad
\bar{z}_A {\cal D}_\mu  z^A = 0 \nonumber \\
 {\cal D}_\mu  \eta _i &=& D_\mu  \eta _i - \frac{1}{4} (\bar{z} \partial
_\mu  z) \eta _i
\end{eqnarray}
Here and further we drop, wherever possible,  the repeated  indices.
In this notations the Lagrangian has the form:
\begin{eqnarray}
 L &=& - \frac{1}{2} R + \frac{i}{2} \varepsilon ^{\mu \nu \rho \sigma }
\bar{\Psi }_\mu {}^i \gamma _5 \gamma _\nu
{\cal D}_\rho  \Psi _{\sigma i} + \frac{1}{2} \bar{\Omega }^i \hat{\cal D}
\Omega _i +
\frac{1}{2} {\cal D}_\mu  \bar{z} {\cal D}_\mu  z - \nonumber \\
  && - \frac{1}{4} A_{\mu \nu }{}^2 + \frac{1}{4} \left\{ \frac{1}{z^2} (z
A_{\mu \nu }) (z(A_{\mu \nu } + i \tilde{A}_{\mu \nu })) + h.c. \right\} -
\nonumber \\
  && - \frac{1}{2} \varepsilon ^{ij} \bar{\Psi }_{\mu i}
\frac{\bar{z}_A}{\bar{z}^2}
(A^{\mu \nu } - \gamma _5 \tilde{A}^{\mu \nu })^A \Psi _{\nu j} -
\frac{1}{2} \varepsilon ^{ij}
\bar{\Omega }_i{}^A \gamma ^\mu  \gamma ^\nu  {\cal D}_\nu  z^A \Psi _{\mu
j} + \nonumber \\
  && + \frac{i}{4} \bar{\Omega }_i{}^A \gamma ^\mu  \left[ (\sigma A)^A -
\frac{z^A(z(\sigma A))}
{z^2} \right] \Psi _{\mu i}
\end{eqnarray}
where now and further on we omit all four fermion terms which will not
be essential for our consideration. This Lagrangian is invariant under
the following local supertransformations:
\begin{eqnarray}
 \delta  e_{\mu r} &=& i (\bar{\Psi }_\mu {}^i \gamma _r \eta _i)
\nonumber \\
 \delta  \Psi _{\mu i} &=& 2 {\cal D}_\mu  \eta _i  - \frac{i}{2}
\varepsilon _{ij}
\frac{(z(\sigma A))}{z^2} \gamma _\mu  \eta ^j \nonumber \\
 \delta  A_\mu {}^A &=& \varepsilon ^{ij} (\bar{\Psi }_{\mu i} z^A \eta
_j) + i (\bar{\Omega }^{iA} \gamma _\mu
\eta _i) \\
 \delta  \Omega _{iA} &=& - \frac{1}{2} \left[ (\sigma A)^A - \frac{z^A
(z(\sigma A))}{z^2}
\right] \eta _i - i \varepsilon _{ij} \hat{\cal D} z^A \eta ^j \nonumber
\\
 \delta  x^A &=& \varepsilon ^{ij} (\bar{\Omega }_{iA} \eta _j) \qquad
\delta  y^A = \varepsilon ^{ij}
(\bar{\Omega }_{iA} \gamma _5 \eta _j) \nonumber
\end{eqnarray}

   Now let us discard for a moment vector multiplets with $a=2,3...m$
and consider the most general interaction of one vector multiplet with
$N=2$ supergravity keeping the scalar field geometry fixed (it is just
$SU(1,1)/U(1)$ in this case). The requirement of axial $U(1)$
invariance and the constraints (1) lead to the following ansatz for
the supertransformations:
\begin{eqnarray}
 \delta \Psi _{\mu i} &=& 2 {\cal D}_\mu  \eta _i - \frac i 2 \varepsilon
_{ij} E_\alpha  (\sigma A)^\alpha  \gamma _\mu  \eta ^j
\nonumber \\
 \delta A_\mu {}^\alpha  &=& \varepsilon ^{ij}(\bar{\Psi }_{\mu i} z^a
K_a{}^\alpha  \eta _j ) - i (\bar{\lambda }^{ia}
\gamma _\mu  K_a{}^\alpha  \eta _i ) \label{h1} \\
 \delta \lambda _{ia} &=& \frac 1 2 \varepsilon _{ab} z^b (M_\alpha
(\sigma A)^\alpha ) \eta _i - i \varepsilon _{ij}
\hat{\cal D} z^a \eta ^j \nonumber \\
 \delta x_a &=& \varepsilon ^{ij} (\bar{\lambda }_{ia} \eta _j ) \quad
\delta y_a = \varepsilon ^{ij} (\bar{\lambda }_{ia}
\gamma _5 \eta _j ) \nonumber
\end{eqnarray}
where $K_a{}^\alpha $ is a constant two by two matrix, and $E_\alpha $ and
$M_\alpha $
are functions of $z^a$ with axial charge $+2$. The requirement of the
closure of the superalgebra leads to the following equation for them:
\begin{equation}
 K_\alpha  E_\beta  + N_\alpha  M_\beta  = \delta _{\alpha \beta }
\end{equation}
where $K_\alpha  = z^a K_a{}^\alpha $, $N_\alpha  = z^a N_{a\alpha }$ and
$N_{a\alpha } = \varepsilon _{ab}
\bar{K}^b{}_\alpha $. This equation allows one to express the functions
$E_\alpha $ and $M_\alpha $ in terms of $K_\alpha $, namely
\begin{equation}
 E_\alpha  = \frac{\varepsilon _{\alpha \beta } N_\beta } \Delta , \qquad
M_\alpha  = - \frac{\varepsilon _{\alpha \beta } K_\beta } \Delta , \qquad
\Delta  = \varepsilon ^{\alpha \beta } K_\alpha  N_\beta
\end{equation}
The corresponding Lagrangian has the form:
\begin{eqnarray}
 L &=& - \frac 1 2 R + \frac i 2 \varepsilon ^{\mu \nu \rho \sigma }
\bar{\Psi }_{\mu i} \gamma _5 \gamma _\nu  {\cal
D}_\alpha  \Psi _\beta  + \frac i 2 \bar{\lambda }^i \gamma ^\mu  {\cal
D}_\mu  \lambda _i + \frac 1 2 {\cal
D}^\mu  z^a {\cal D}_\mu  \bar{z}_a - \nonumber \\
 && - \frac 1 4 [ f_{\alpha \beta } A_{\mu \nu }{}^\alpha  A_{\mu \nu
}{}^\beta  - g_{\alpha \beta } A_{\mu \nu }{}^\alpha
\tilde{A}_{\mu \nu }{}^\beta ] - \frac 1 2 \varepsilon ^{ij} \bar{\lambda
}_{ia} \gamma ^\mu  \gamma ^\nu  {\cal D}_\nu
z^a \Psi _{\mu j} \nonumber \\
 && - \frac 1 2 \varepsilon ^{ij} \bar{\Psi }_{\mu i} (A^{\mu \nu } -
\gamma _5 \tilde{A }^{\mu \nu })^\alpha
\bar{E}^\alpha  \Psi _{\nu j} - \frac i 4 \bar{\lambda }^i{}_a \varepsilon
^{ab} \bar{z_b} \gamma ^\mu
M_\alpha (\sigma A)^\alpha  \Psi _{\mu i} \label{h2}
\end{eqnarray}
where $f_{\alpha \beta }$ and $g_{\alpha \beta }$ are real and symmetric
functions to be
determined. The requirement of the Lagrangian to be invariant under
the supertransformations given above leads to the following equations:
\begin{eqnarray}
 f_{\alpha \beta } = 2 (E_\alpha  \bar{E}_\beta  + M_\alpha  \bar{M}_\beta
) && \nonumber \\
 \frac{\delta (f_{\alpha \beta } + \gamma _5 g_{\alpha \beta })}{\delta
z^a} = 4 \varepsilon _{ab} z^b M_{(\alpha } E_{\beta )} &&
\label{eq2}
\end{eqnarray}
The first one determines the function $f_{\alpha \beta }$ and, in order
this
function to be symmetric, leads to the (only) constraint on the matrix
$K_a{}^\alpha $:
\begin{equation}
 \varepsilon ^{\alpha \beta } \bar{K}^a{}_\alpha  K_{a\beta } = 0
\label{c1}
\end{equation}
As for the second one, it could be solved by the following ansatz:
\begin{equation}
 f_{\alpha \beta } + \gamma _5 g_{\alpha \beta } = H_\alpha  E_\beta  +
G_\alpha  M_\beta
\end{equation}
where $H_\alpha  = z^a H_{a\alpha }$, $G_\alpha  = z^a G_{a\alpha }$ and
$G_{a\alpha } = - \varepsilon _{ab}
\bar{H}^b{}_\alpha $. Now both equations (\ref{eq2}) are satisfied if
\begin{equation}
 \bar{H}^a{}_\alpha  K_{a\beta } + H_{a\alpha } \bar{K}^a{}_\beta  = - 2
\delta _{\alpha \beta }
\end{equation}
Note, that this equation determines the matrix $H_{a\alpha }$ only up to
the arbitrary constant due to the possibility to make a shift $H_{a\alpha
}
\to  H_{a\alpha } + \omega  \gamma _5 \varepsilon _{\alpha \beta }
K_a{}^\beta $. This, in turn, implies that the
function $g_{\alpha \beta }$ is determined up to the arbitrary constant,
but
that leads to the Lagrangians that differ only by the total
divergency.

   Thus, we have managed to construct a whole class of Lagrangians
with the same scalar field geometry and different dual forms for
vector fields, which are determined by the constant matrix $K_a{}^\alpha $
with the only constraint (\ref{c1}) on it. As partial cases we have
usual ''minimal'' case, corresponding to the choice $K_a{}^\alpha  =
\left(\begin{array}{cc} 1 & 1 \\ -1 & 1 \end{array}\right)$, and the
dual version that we mentioned above, with $K_a{}^\alpha  = \left(
\begin{array}{cc} 1 & \gamma _5 \\ -1 & - \gamma _5 \end{array}\right)$
(we will
give corresponding Lagrangian and the supertransformations in the next
section).

   At thirst sight it seems that it is not easy to generalize the
solution obtained to the case of arbitrary number of vector
multiplets. In order to do that note, first of all, that the part of
the supertransformations for gravitini containing vector fields could
be rewritten in a very suggestive form:
\begin{equation}
 \delta  \Psi _{\mu i} = - \frac{i}{2} \varepsilon _{ij} \frac{(z^a
\tilde{E}_{a\alpha }
(\sigma A)^\alpha )}{z^a g_{ab} z^b} \gamma _\mu  \eta ^j \nonumber
\end{equation}
where we introduced:
\begin{equation}
 \tilde{E}_{a\alpha } = \varepsilon _{ab} \varepsilon _{\alpha \beta }
\bar{K}^{b\beta } \qquad g_{ab} = \varepsilon ^{\alpha \beta }
K_{a\alpha } N_{b\beta } \nonumber
\end{equation}
Moreover, in the case when $det(g_{ab}) \ne  0$ the corresponding part
of the $\lambda $-supertransformations could be rewritten as:
\begin{equation}
 \delta  \lambda _{ia} = - \frac{1}{2}\left[ E_{a\alpha } (\sigma
A)^\alpha  - \frac{g_{ab} z^b (z^c
E_{c\alpha } (\sigma A)^\alpha )}{(z g z)} \right] \eta _i \nonumber
\end{equation}

   Now we are ready to consider the generalization of this formulas on
the case of arbitrary number of vector multiplets. For that we make
the following ansatz for the supertransformations:
\begin{eqnarray}
 \delta  e_{\mu r} &=& i (\bar{\Psi }_\mu {}^i \gamma _r \eta _i)
\nonumber \\
 \delta  \Psi _{\mu i} &=& 2 {\cal D}_\mu  \eta _i - \frac{i}{2}
\varepsilon _{ij} \frac{z^A E_{A\alpha }
(\sigma A)^\alpha }{(z g z)} \gamma _\mu  \eta ^j \nonumber \\
 \delta  A_m{}^\alpha  &=& \varepsilon ^{ij} (\bar{\Psi }_{\mu i} z^A
K_A{}^\alpha  \eta _j) + i (\bar{\Omega }^{iA}
\gamma _\mu  K_A{}^\alpha  \eta _i) \label{g1} \\
 \delta  \Omega _{iA} &=& - \frac{1}{2} \left[ E_{A\alpha } (\sigma
A)^\alpha  - \frac{g_{AB} z^B
(z^C E_{C\alpha } (\sigma A)^\alpha )}{(z g z)} \right] \eta _i - i
\varepsilon _{ij} \hat{\cal D} z^A
\eta ^j \nonumber \\
 \delta  x^A &=& \varepsilon ^{ij} (\bar{\Omega }_{iA} \eta _j) \qquad
\delta  y^A = \varepsilon ^{ij}
(\bar{\Omega }_{iA} \gamma _5 \eta _j) \nonumber
\end{eqnarray}
and for the Lagrangian:
\begin{eqnarray}
 L &=& - \frac{1}{2} R + \frac{i}{2} \varepsilon ^{\mu \nu \rho \sigma }
\bar{\Psi }_\mu {}^i \gamma _5 \gamma _\nu
{\cal D}_\rho  \Psi _{\sigma i} + \frac{i}{2} \bar{\Omega }^{iA} \hat{\cal
D} \Omega _{iA} +
\frac{1}{2} {\cal D}_\mu  \bar{z}_A {\cal D}_\mu  z^A - \nonumber \\
  && - \frac{1}{4} A_{\mu \nu }{}^\alpha  E_{A\alpha } \bar{E}^A{}_\beta
A_{\mu \nu }{}^\beta  +
\frac{1}{4} \left[ \frac{(z E A_{\mu \nu })(z E (A_{\mu \nu } + \gamma _5
\tilde{A}_{\mu \nu }))}{(z g z)} + h.c. \right] - \nonumber \\
  && - \frac{1}{2} \varepsilon ^{ij} \bar{\Psi }_{\mu i} \frac{\bar{z}_A
\bar{E}^A{}_\alpha
(A^{\mu \nu } - \gamma _5 \tilde{A}^{\mu \nu })^\alpha }{(\bar{z} \bar{g}
\bar{z})} \Psi _{\nu j} -
\frac{1}{2} \varepsilon ^{ij} \bar{\Omega }^{iA} \gamma ^\mu  \gamma ^\nu
{\cal D}_\nu  z^A \Psi _{\mu j} +
\nonumber \\
  && + \frac{i}{4} \bar{\Omega }^{iA} \gamma ^\mu  \left[ E_{A\alpha }
(\sigma A)^\alpha  -
\frac{g_{AB}z^B (z E (\sigma A))}{(z g z)} \right] \Psi _{\mu i}
\label{g2}
\end{eqnarray}
In this, the requirements of the closure of the superalgebra and the
invariance of the Lagrangian give:
\begin{equation}
 \bar{K}^{A\alpha } E_{A\beta } = \delta ^\alpha {}_\beta  \qquad g_{AB} =
K_A{}^\alpha  E_{B\alpha } \qquad
E_{A[\alpha } \bar{E}^A{}_{\beta ]} = 0
\end{equation}
The first two equations allow one to express $K_A{}^\alpha $ and $g_{AB}$
in
terms of the $E_{A\alpha }$ while the last one turns out to be the only
constraint on $E_{A\alpha }$. Note that in such model vector fields
$A_\mu {}^\alpha $ and spinor $\Omega _{iA}$ and scalar $z^A$ fields carry
different kind of indices exactly as in the general construction of
\cite{D'A91}.

\section{Spontaneous supersymmetry breaking}

   In this section we would like to explain how the results obtained
above are connected with the problem of spontaneous supersymmetry
breaking. As we have already mentioned there exist the hidden sector
($N=2$ supergravity, vector and hypermultiplet) admitting spontaneous
supersymmetry breaking with two arbitrary scales and without a
cosmological term. The two vector fields (graviphoton and the matter
one) enter through complex combinations $A_\mu  \pm  \gamma _5 B_\mu $ and
it seemed
that the only natural generalization of this model to the case of
arbitrary number of vector multiplets is the model with the scalar
field geometry $SO(2,m)/SO(m)\otimes SO(2)$. Indeed it has been shown
\cite{Zin86a,Zin90} that for this case one can have spontaneous
supersymmetry breaking for arbitrary number of vector and
hypermultiplets and the corresponding soft breaking terms have been
calculated.

   Now using the general formulas given in the previous section we can
obtain as a partial case the generalization of our hidden sector to
the case of scalar field geometry being $SU(1,m)/SU(m)\otimes U(1)$. But
the
case of interest turns to be the singular one corresponding to the
degenerate matrix $E_{a\alpha }$ and hence $g_{ab}$. Let us stress however
that our general solution for one vector multiplets had no
singularities in this case. It is in rewriting it in a "covariant"
form that one faces the requirement the matrix $g_{ab}$ to be
nondegenerate. So let us split the vector fields onto the hidden
sector (graviphoton and one vector fields) and ''matter'' ones.
Moreover, in order to be able to compare the results with the
previously obtained, it turns out to be useful to write all the
formulas in term of the physical scalar and spinor fields. In order to
solve the constraints (\ref{e1}) we introduce a kind of "light cone"
variables, namely, we split $z^A = \{x+y, x-y, z^a\}$, $a = 2,3....m$.
Then the scalar field constrains takes the form:
\begin{equation}
 \bar{x} y + x \bar{y} = 1 + \frac{\bar{z}_a z^a}{2}
\end{equation}
Now we introduce a new variable $\pi  = -i(\frac{\bar{x}}{\bar{y}} -
\frac{x}{y})$ (a choice of such combination will become clear later)
so that we can exclude the field $x$ and reexpress all formulas in
term of $y$, $\pi $ and $z^a$. It turns out that in order to have
canonical form of the scalar field kinetic terms one has to make a
change $z^a \to  y z^a$. Then the only scalar field having nonzero axial
charge is $y$ and we can use local axial $U(1)$ invariance to make it
real. Denoting this real field by $\Phi $ we obtain:
\begin{equation}
 \frac{1}{2} {\cal D}_\mu  \bar{z}_A {\cal D}_\mu  z^A = \left(
\frac{\partial _\mu
\Phi }{\Phi } \right)^2 + \frac{1}{2} \Phi ^2 \partial _\mu  \bar{z}
\partial _\mu  z - \frac{1}{4}
U_\mu {}^2
\end{equation}
where $U_\mu  = (\bar{z}_A \partial _\mu  z^A) = \Phi ^2 (i \partial _\mu
\pi  + \frac{1}{2}
(\bar{z}\stackrel{\leftrightarrow }\partial  z))$.

   Analogously, let us introduce $\Omega _{iA} = \{ -(\lambda _i + \chi
_i), (\chi _i -
\lambda _i), \Omega _{ia} \}$. Then the spinor field constraint
\begin{equation}
 x \lambda _i + y \chi _i - \frac{z^a}{2} \Omega _{ia} = 0
\end{equation}
allows one to exclude spinor $\chi _i$. Again in order to have canonical
kinetic terms one has to make two changes $\Omega _{ia} \to  \Omega _{ia}
+ \bar{z}_a
\lambda _i$ and $\lambda _i \to  \frac{\bar{y}}{\sqrt{2}} \lambda _i$. We
have
\begin{equation}
 \frac{i}{2} \bar{\Omega }^{iA} \hat{\cal D} \Omega _{iA} = \frac{i}{2}
\bar{\lambda }^i \gamma ^\mu  [D_\mu  - \frac{3}{4} U_\mu  ] \lambda _i +
\frac{i}{2} \bar{\Omega }^{ia}
\gamma ^\mu  [D_\mu  - \frac{1}{4} U_\mu  ] \Omega _{ia} -
\frac{i}{\sqrt{2}} \bar{\lambda }^i \gamma ^\mu
\Phi  \partial _\mu  z^a \Omega _{ia}
\end{equation}

   It is not hard to rewrite supertransformations in term of the new
fields (here we give the terms without vector fields only):
\begin{eqnarray}
  \delta  e_{\mu r} &=& i (\bar{\Psi }_\mu {}^i \gamma _r \eta _i) \qquad
 \delta  \Psi _{\mu i} = 2 D_\mu  \eta _i -  \frac{1}{2} U_\mu  \eta _i
\nonumber \\
 \delta  \lambda _i &=& - i \varepsilon _{ij} \gamma ^\mu  \left[ \sqrt{2}
\frac{\partial _\mu  \Phi }{\Phi } +
\frac{1}{\sqrt{2}} U_\mu  \right] \eta ^j \qquad \delta  \Omega _{ia} = -
i \varepsilon _{ij} \gamma ^\mu  \Phi
\partial _\mu  z^a \eta ^j \nonumber \\
 \delta  \Phi  &=& \frac{\Phi }{\sqrt{2}} \varepsilon ^{ij} (\bar{\lambda
}_i \eta _j) \qquad \delta  \pi  =
\frac{\sqrt{2}}{\Phi ^2} \varepsilon ^{ij}(\bar{\lambda }_i \gamma _5 \eta
_j) - \frac{1}{\Phi } \varepsilon ^{ij}
(\bar{\Omega }_{ia} \gamma _5 z^a \eta _j) \\
 \delta  x^a &=& \frac{1}{\Phi } \varepsilon ^{ij} (\bar{\Omega }_{ia}
\eta _j) \qquad \delta  y^a =
\frac{1}{\Phi } \varepsilon ^{ij} (\bar{\Omega }_{ia} \gamma _5 \eta _j)
\nonumber
\end{eqnarray}
The corresponding part of the Lagrangian looks like:
\begin{eqnarray}
 L &=& - \frac{1}{2} R + \frac{i}{2} \varepsilon ^{\mu \nu \rho \sigma }
\bar{\Psi }_\mu {}^i \gamma _5 \gamma _\nu
[D_\rho  - \frac{1}{4} U_\rho  ] \Psi _{\sigma i} + \frac{i}{2}
\bar{\lambda }^i \gamma ^\mu  [D_\mu  -
\frac{3}{4} U_\mu  ] \lambda _i + \nonumber \\
  && + \frac{i}{2} \bar{\Omega }^{ia} \gamma ^\mu  [D_\mu  - \frac{1}{4}
U_\mu  ] \Omega _{ia}
+ \left( \frac{\partial _\mu  \Phi }{\Phi } \right)^2 - \frac{1}{4} U_\mu
{}^2 +
\frac{1}{2} \Phi ^2 \partial _\mu  \bar{z} \partial _\mu  z - \nonumber \\
  && - \frac{1}{2} \varepsilon ^{ij} \bar{\lambda }_i \gamma ^\mu  \gamma
^\nu  [\sqrt{2} \frac{\partial _\nu  \Phi }{\Phi }
+ \frac{1}{\sqrt{2}} U_\nu  ] \Psi _{\mu j} - \frac{1}{2} \varepsilon
^{ij} \bar{\Omega }_i{}^a
\gamma ^\mu  \gamma ^\nu  \Phi  \partial _\nu  z^a \Psi _{\mu j} -
\nonumber \\
 && - \frac{i}{\sqrt{2}} \bar{\lambda }^i \gamma ^\mu  \Phi  \partial _\mu
 z^a \Omega _i{}^a
\end{eqnarray}
One can see that the $\pi $ field enter through the divergency $\partial
_\mu  \pi $
only, one of the isometries acting as a translation $\pi  \to  \pi  +
\Lambda $. That
explains our special choice for $\pi $. Let us stress, that the last two
formulas are universal in a sense that they are determined by the
choice of scalar field geometry and do not depend on the choice of
dual version.

   Now it is not hard to get from the formulas (\ref{h1}), (\ref{h2})
our hidden sector in terms of the physical fields (note that our
current notations differ a little from ones used in \cite{Zin86}).
Additional terms for the supertransformations looks like:
\begin{eqnarray}
 \delta  \Psi _{\mu i} &=& - \frac{i}{4\Phi } \varepsilon _{ij} \sigma (A
- \gamma _5 B) \gamma _\mu  \eta ^j \nonumber \\
 \delta  A_\mu  &=& \Phi  \left[ \varepsilon ^{ij} (\bar{\Psi }_{\mu i}
\eta _j) + \frac{i}{\sqrt{2}}
(\bar{\lambda }^i \gamma _\mu  \eta _i) \right] \nonumber \\
 \delta  B_\mu  &=& \Phi  \left[ \varepsilon ^{ij} (\bar{\Phi }_{\mu i}
\gamma _5 \eta _j) + \frac{i}{\sqrt{2}}
(\bar{\lambda }^i \gamma _\mu  \gamma _5 \eta _i) \right] \\
 \delta  \lambda _i &=& - \frac{1}{2\sqrt{2}\Phi } \sigma (A + \gamma _5
B) \eta _i \nonumber
\end{eqnarray}
and whose for the Lagrangian have the form:
\begin{eqnarray}
 L &=& - \frac{1}{4\Phi ^2} (A_{\mu \nu }{}^2 + B_{\mu \nu }{}^2) -
\frac{\pi }{4}
(A_{\mu \nu } \tilde{A}_{\mu \nu } + B_{\mu \nu } \tilde{B}_{\mu \nu }) -
\nonumber \\
  && - \frac{1}{4\Phi } \varepsilon ^{ij} \bar{\Psi }_{\mu i} (A^{\mu \nu
} - \gamma _5 \tilde{A}^{\mu \nu } +
\gamma _5 B^{\mu \nu } + \tilde{B}^{\mu \nu }) \Psi _{\nu j} + \nonumber
\\
  && + \frac{i}{4\sqrt{2} \Phi } \bar{\lambda }^i \gamma ^\mu  \sigma (A +
\gamma _5 B) \Psi _{\mu i}
\end{eqnarray}

  Note here that for this model (just because the matrix $g_{ab}$ is
degenerate) the global symmetry of the Lagrangian is enhanced. At
first we have axial $U(1)$ transformations acting on the spinor fields
as well as on the complex vector field $A_\mu  + \gamma _5 B_\mu $. Then
we have
scale transformations such as $\Phi  \to  e^\Lambda  \Phi $, $(A_\mu
,B_\mu ) \to  e^\Lambda (A_\mu ,B_\mu )$,
$\pi  \to  e^{-2\Lambda } \pi $. And at last the whole Lagrangian
including terms with
vector fields is invariant under translations $\pi  \to  \pi  + \Lambda $.

   As for the matter vector fields, there are two ways. One can use
general formulas (\ref{g1}), (\ref{g2}) and, making some kind of
regularization for matrix $K_a{}^\alpha $ in order to avoid singularities,
rewrite them in term of the physical fields (keeping only terms which
survive when regularization parameter goes to zero and making field
rescaling if necessary). On the other hand one can start directly
from the hidden sector and add to it the additional vector multiplets.
We have explicitly checked that both ways lead to the same answer,
namely, that
\begin{eqnarray}
 \delta  C_\mu {}^a &=& i (\bar{\Omega }^{ia} \gamma _\mu  \eta _i) +
\varepsilon ^{ij} (\bar{\Psi }_{\mu i} \Phi  z^a
\eta _j) + \frac{1}{\sqrt{2}} (\bar{\lambda }^i \gamma _\mu  \Phi  z^a
\eta _i) \nonumber \\
 \delta  \Omega _{ia} &=& - \frac{1}{2} [ (\sigma C)^a - x^a (\sigma A) -
y^a (\sigma B) ] \eta _i
\end{eqnarray}
are the only additional terms for the supertransformations, while the
Lagrangian has to be completed with:
\begin{eqnarray}
 L &=& - \frac{1}{4} (C_{\mu \nu }{}^a)^2 + \frac{1}{2} C_{\mu \nu }{}^a [
(x^a
A_{\mu \nu } + y^a B_{\mu \nu }) - ( y^a \tilde{A}_{\mu \nu } - x^a
\tilde{B}_{\mu \nu }) ] -
\nonumber \\
  && - \frac{1}{4} (x^a A_{\mu \nu } + y^a B_{\mu \nu }) [ (x^a A_{\mu \nu
} + y^a
B_{\mu \nu }) - (y^a \tilde{A}_{\mu \nu } - x^a \tilde{B}_{\mu \nu }) ] +
\nonumber \\
  && + \frac{i}{4} \bar{\Omega }^{ia} \gamma ^\mu  [ (\sigma C)^a - x^a
(\sigma A) - y^a (\sigma B) ]
\Psi _{\mu i}
\end{eqnarray}
   As we have already mentioned the hidden sector \cite{Zin86}
includes also the hypermultiplet in a formulation with a singlet
$\tilde{\Phi }$ and a triplet $\pi _i{}^j$ of scalar fields as well as a
doublet $\chi _i$ of spinor fields. It is not hard to introduce such a
multiplet in our model by adding to the Lagrangian
\begin{eqnarray}
 L &=& \frac{1}{2} \left( \frac{\partial _\mu  \tilde{\Phi }}{\tilde{\Phi
}} \right) +
\frac{\tilde{\Phi }^2}{2} (\partial _\mu  \vec{\pi })^2 - \frac{1}{2}
\varepsilon ^{ij} \bar{\chi }_i
\gamma ^\mu  \gamma ^\nu  [ \delta _j{}^k \frac{\partial _\nu  \tilde{\Phi
}}{\tilde{\Phi }} + \tilde{\Phi } \partial _\nu
\pi _j{}^k ] \Psi _{\mu k} + \nonumber \\
  && + \frac{i}{2} \bar{\chi }^i \gamma ^\mu  [D_\mu  + \frac{1}{4} U_\mu
] +
\frac{1}{8\Phi } \varepsilon ^{ij} \bar{\chi }_i \sigma (A - \gamma _5 B)
\chi _j - \nonumber \\
  && - \frac{i}{4} \tilde{\Phi } [ \varepsilon ^{\mu \nu \rho \sigma }
\bar{\Psi }_\mu {}^i \gamma _5 \gamma _\nu  \partial _\rho
\pi _i{}^j \Psi _{\sigma j} + \bar{\lambda }^i \gamma ^\mu  \partial _\mu
\pi _i{}^j \lambda _j - \nonumber \\
  && \qquad - \bar{\chi }^i \gamma ^\mu  \partial _\mu  \pi _i{}^j \chi _j
+ \bar{\Omega }^{ia} \gamma ^\mu  \partial _\mu
\pi _i{}^j \Omega _i{}^a ]
\end{eqnarray}
and to the supertransformations:
\begin{eqnarray}
 \delta ' \Psi _{\mu i} &=& - \tilde{\Phi } \partial _\mu  \pi _i{}^j \eta
_j \nonumber \\
 \delta  \chi _i &=& - i \varepsilon _{ij} \gamma ^\mu  [ \delta _j{}^k
\frac{\partial _\mu  \tilde{\Phi }}{\tilde{\Phi }} +
\tilde{\Phi } \partial _\mu  \pi _j{}^k ] \eta _k \\
 \delta  \tilde{\Phi } &=& \tilde{\Phi } \varepsilon ^{ij} (\bar{\chi }_i
\eta _j) \qquad \delta  \vec{\pi } =
\frac{1}{\tilde{\Phi }} \varepsilon ^{ij} (\bar{\chi }_i (\vec{\tau
})_j{}^k \eta _k) \nonumber
\end{eqnarray}
Here $(\vec{\tau })_i{}^j$ - is the usual Pauli matrices and we use
$\pi _i{}^j = \vec{\pi } (\vec{\tau })_i{}^j$.

   The peculiar feature of this formulation for the hypermultiplets is
the invariance of the Lagrangian under translations $\vec{\pi } \to
\vec{\pi }
+ \vec{\Lambda }$. This allows one to introduce masses for two vector
fields
$A_\mu $ and $B_\mu $ using e.g. fields $\pi _1$ and $\pi _2$ as the
Goldstone
ones. So let us make in the Lagrangian and in the supertransformations
the replacement:
\begin{equation}
 \partial _\mu  \pi _i{}^j \to  \partial _\mu  \pi _i{}^j - A_\mu
(t_1)_i{}^j - B_\mu  (t_2)_i{}^j
\end{equation}
where $(t_{1,2})_i{}^j = m_{1,2} (\tau _{1,2})_i{}^j$.
As usual in supergravity such a replacement spoils the invariance of
the Lagrangian under the supertransformations but it can easily be
restored by adding to the lagrangian
\begin{eqnarray}
 \Delta L &=& \Phi  \tilde{\Phi } \left\{ - \frac{1}{4} \bar{\Psi }_{\mu
i} \sigma ^{\mu \nu } \varepsilon ^{ij}
(M_+)_j{}^k \Psi _{\nu k} - \frac{i}{2\sqrt{2}} \bar{\Psi }_\mu {}^i
\gamma ^\mu  (M_-)_i{}^j
[ \lambda _j + \sqrt{2} \chi _j ] - \right. \nonumber \\
   && \left. \qquad - \frac{1}{\sqrt{2}} \bar{\lambda }_i \varepsilon
^{ij} (M_-)_j{}^k
\chi _k - \frac{1}{4} \bar{\chi }_i \varepsilon ^{ij} (M_-)_j{}^k \chi _k
\right\}
\end{eqnarray}
where $(M_{\pm })_i{}^j = (t_1)_i{}^j \pm  \gamma _5 (t_2)_i{}^j$ and to
the
supertransformations:
\begin{eqnarray}
 \delta ' \Psi _{\mu i} &=& \frac{i}{2} \Phi  \tilde{\Phi } \gamma _\mu
\varepsilon ^{ij} (M_+)_j{}^k \eta _k
\nonumber \\
 \delta ' \lambda _i &=& \frac{1}{\sqrt{2}} \Phi  \tilde{\Phi }
(M_+)_i{}^j \eta _j \\
 \delta ' \chi _i &=& \Phi  \tilde{\Phi } (M_+)_i{}^j \eta _j \nonumber
\end{eqnarray}

   From this formulas one can see that in this model we can really
have spontaneous supersymmetry breaking with two different scales,
gravitini masses being connected with the vector field masses through
the relations $\mu _1 = \frac{m_1 - m_2}{2}$ and $\mu _2 = \frac{m_1 +
m_2}{2}$. There are no any scalar field potential in this model so the
cosmological term is automatically equals to zero. Note that one
could introduce a gauge interaction for vector multiplets for some
nonabelian gauge group $G$ as well. In this case a nonzero scalar field
potential for the $z^a$ fields will be generated but the cosmological
term remains to be zero. At the same time one can see that no soft
breaking terms are generated in this model in great contrast with the
case of the geometry $SO(2,m)/SO(m)\otimes SO(2)$.

\section{Conclusion}

   Thus we have shown that for the system of $N=2$ supergravity -
vector multiplets with fixed scalar field geometry one can have a
whole family of the Lagrangians interpolating between previously known
dual versions for such models. We have seen that it is the choice of
dual version that determines the possibility to have spontaneous
supersymmetry breaking without a cosmological term while the scalar
field geometry determines the pattern of soft breaking terms that are
generated (or not generated) after symmetry breaking takes place.

\vspace{0.3in}
{\large \bf Acknowledgments}
\vspace{0.2in}

   Work supported by International Science Foundation grant RMP000 and
by Russian Foundation for Fundamental Research grant 94-02-03552.

\end{document}